\documentclass[reprint,aps,pre,floatfix,twocolumn,showpacs,superscriptaddress
]{revtex4-1}

\usepackage{times}
\usepackage[pdftex]{graphicx}
\usepackage{latexsym,amsmath}
\usepackage[english]{babel}
\usepackage[pdftex]{color}
\usepackage[dvipsnames]{xcolor}
\usepackage{microtype}
\usepackage{comment}
\usepackage[utf8]{inputenc}
\usepackage[normalem]{ulem}
\usepackage{etoolbox}
\usepackage[justification=centerlast,labelfont=bf,font=small]{caption}

\usepackage{latexsym,amsmath}
\usepackage{amsbsy}
\usepackage{amssymb}
\usepackage{epstopdf}
\usepackage{comment}
\usepackage{multirow,tabularx}
 \usepackage[normalem]{ulem}

 \usepackage[hidelinks,unicode=true]{hyperref}
 \hypersetup{colorlinks=true,
 	linkcolor=blue,
 	urlcolor=blue,
 	citecolor=blue,
 	pdfhighlight=/N
 }

\newcommand{\av}[1]{\langle #1 \rangle}

\newcommand{\FigPath}{./}

\begin{document}

\title{Quantifying echo chamber effects in information spreading over political
  communication networks}

\author{Wesley Cota}
\affiliation{Departamento de F\'{\i}sica, Universidade Federal de  Vi\c{c}osa,
  36570-900 Vi\c{c}osa, Minas Gerais, Brazil}

\author{Silvio C. Ferreira}
\affiliation{Departamento de F\'{\i}sica, Universidade Federal de Vi\c{c}osa,
  36570-900 Vi\c{c}osa, Minas Gerais, Brazil}
\affiliation{National Institute of Science and Technology for Complex Systems, Brazil}

\author{Romualdo Pastor-Satorras} \affiliation{Departament de F\'{\i}sica,
  Universitat Polit\`ecnica de Catalunya, Campus Nord B4, 08034 Barcelona,
  Spain}

\author{Michele Starnini} \email{Corresponding author:
  michele.starnini@gmail.com} \affiliation{ISI Foundation, via Chisola 5, 10126
  Torino, Italy}

\begin{abstract}
  Echo chambers in online social networks, in which users prefer to interact
  only with ideologically-aligned peers, are believed to facilitate
  misinformation spreading and contribute to radicalize political discourse. In
  this paper, we gauge the effects of echo chambers in information spreading
  phenomena over political communication networks. Mining 12 million Twitter
  messages, we reconstruct a network in which users interchange opinions related
  to the impeachment of the former Brazilian President Dilma Rousseff. We define
  a continuous {political position} parameter,
  independent of the network's structure, that allows to quantify the presence
  of echo chambers in the strongly connected component of the network, reflected
  in two well-separated communities of similar sizes with opposite views of the
  impeachment process. By means of simple spreading models, we show that the
  capability of users in propagating the content they produce, measured by the
  associated spreadability, strongly depends on their attitude. Users expressing
  pro-impeachment sentiments are capable to transmit information, on average, to
  a larger audience than users expressing anti-impeachment sentiments. 
  Furthermore, the users' spreadability is correlated to the diversity, in terms
  of political position, of the audience reached. Our method can be
  exploited to identify the presence of echo chambers and their effects across
  different contexts and shed light upon the mechanisms allowing to break echo
  chambers.
\end{abstract}


\maketitle

Online social networks in which users can be both consumers and producers of
content, such as Twitter or Facebook, provide means to exchange information in
an almost instantaneous, inexpensive, and not mediated way, forming a substrate
for the spread of information with unprecedented capabilities.  These new
channels of communication have enormously altered the way in which we take
decisions, form political opinions, align in front of different issues, or
choose between the adoption of different technological
options~\cite{masum2011reputation}.  Such online communication networks are
orders of magnitude larger than those classically available in social
sciences~\cite{wass94}, making it possible to perform measurements and
experiments that have led to the definition of a new \textit{computational social
  science}~\cite{lazer2009life}.

One of the characteristic features of online communication networks is their
marked degree of homophily. That is, individuals prefer to interact with other
individuals who are similar to them, or share the same views and
orientations~\cite{doi:10.1111/jcom.12077,citeulike:11235960,Aral21544}.
Homophily leads to a natural polarization of societies into groups with
different perspectives, that leave digital fingerprints in the online realm, and
provide researchers with large-scale data sets for the study of polarization in
different contexts, such as the US and French presidential
elections~\cite{Hanna:2013:PAP:2508436.2508438}, secular vs. Islamist
discussions during the 2011 Egyptian revolution~\cite{Weber2013,
  Borge-Holthoefer15}, or the 15M movement of 2011 in Spain~\cite{6239}.
Political orientation, in particular, has been shown to drive the segregation
of online communication networks into separated
communities~\cite{Truthy_icwsm2011politics,Conover2012}. The presence of these
clusters formed by users with a homogeneous content production and diffusion
have been named \textit{echo chambers}~\cite{garrett2009echo}, referring to the
situation in which one's beliefs are reinforced due to repeated interactions
with individuals sharing the same points of view
\cite{Garimella:2018:PDS:3178876.3186139}. Echo chambers have been shown to
pervade the offline realm \cite{city20893}, to be related to the spreading of
misinformation~\cite{del2016spreading,DBLP:journals/corr/VicarioVBZSCQ16}, or
the development of ideological radicalism~\cite{doi:10.1177/1461444809342775}.
Recent studies, however, have challenged the impact of echo chambers and
partisan segregation in communication networks over online social
media~\cite{doi:10.1177/0956797615594620, doi:10.1080/1369118X.2018.1428656}.

This novel debate calls for a quantitative analysis aimed at identifying the
impact of political position over the diffusion of information. In this paper,
we contribute to this endeavor by proposing a new 
 continuous measure of {political orientation}, and by quantifying its effects in the description
of associated echo chambers and the behavior of information spreading processes
running on top of online communication networks. 
To this aim, we reconstruct a political communication (PC) network, in which individuals exchange messages related to the impeachment process of the former Brazilian President Dilma Rousseff~\cite{wikidilma}, over the social microblogging platform
Twitter. We collected over 12 million tweets from half million users, in a time
window of 9 months, covering the main events related to the impeachment process
and related street protests. The political orientation of users was inferred by
means of a hand-tagged analysis of the hashtags adopted in the messages, which
are assigned anti-impeachment, pro-impeachment, or neutral sentiments.

The topological analysis of the resulting static PC network reveals clusters of
individuals sharing similar opinions, defining the presence of echo
chambers. We gauge the impact of these echo chambers over information spreading
by means of simple spreading models, characterizing the efficiency of single
users to disseminate information, or \textit{spreadability}. Differently from
previous studies, to characterize information diffusion we take into account the
full temporal evolution of the social interactions, representing it in terms of
a temporal network~\cite{Holme:2015,lambiotte2016}, so to ensure that the
spreading process respects the communication dynamics. Our analysis shows that the spreadability of users is strongly correlated with their political orientation: information sent by pro-impeachment individuals spreads throughout the network much better
than messages sent by other users.  Furthermore, by analyzing the composition of
the audience reached, we discover that users with larger spreadability are able
to reach individuals with more diverse sentiments, actually escaping their echo
chamber.

%

\section*{Polarization in political communication networks}
\label{sec:constr-time-vary}

In Twitter, users post real-time short messages (tweets), sometimes annotated
with hashtags indicating the topic of the message, that are broadcast to the
community of their followers.  A user can also transmit (retweet) messages from
other users, forwarding it to its own followers, as a way to endorse its
content. Analysis of retweets (RTs) have been used to study viral propagation of
information in several
contexts~\cite{Galuba:2010:OTP:1863190.1863193,Jenders2013,Ratkiewicz2011}.
However, RTs do not involve an explicit effort of content production and do not
convey a specific communication target.  For this reason, here we discard RTs
form our analysis and focus on tweets that include an explicit mention to
another user, with the purpose of establishing or continuing a discussion on
some topic, carrying even personal messages~\cite{10.1371/journal.pone.0029358}.
This choice allows us to single out only actual social interactions between
users, so as to reconstruct a communication network in which people actually
exchange information, discuss, and form their opinion reacting in real time to
ongoing political events.

As an example of strongly polarized political discussion, we focus on the debate
ensuing the impeachment process of the former Brazilian President Dilma
Rousseff, taking place during 2016. Tweets related to the impeachment process
were gathered by setting a specific filter for tweets containing selected
keywords. The keyword list was kept up to date as new trending topics
continuously appeared on Twitter, see Supplementary Information (SI).
Furthermore, the full dynamics of social interactions was taken into account by
including the real timing of tweets in a temporal network
representation~\cite{lambiotte2016}.  This ensures that information diffusion
over the resulting temporal PC network follows time-respecting paths, which are
expected to have an effect in slowing down or speeding up the spreading
dynamics~\cite{temporalnetworksbook}.  From this temporal network
representation, a static aggregated, directed, weighted
network~\cite{Newman2010} was constructed, in which a directed link from node
$i$ to node $j$ indicates a message sent from user $i$ to user $j$. The
associated weight $W_{ij}$ represents the number of tweets from $i$ to $j$.
Note that, while we keep the temporal evolution of the social interactions when
addressing information spreading dynamics, we measure political position
over the aggregated, static network representation.

Twitter is known to be populated by social bots, that contribute to the
spreading of misinformation and poison political
debate~\cite{Ferrara:2016:RSB:2963119.2818717}. Recent studies revealed that
while bots tend to interact with humans, e.g. by targeting influential users,
the opposite behavior, interactions from humans toward bots, are far less
frequent \cite{Stella12435,Shao2018}.  For this reason, once reconstructed the
aggregated network, we extracted its largest strongly connected component
(SCC)~\cite{Newman2010} to possibly discard social bots and ensure that only
real social interactions between users are considered. Our analysis is
restricted only to the set of individuals composing this SCC.  This choice comes
at the cost of greatly reducing the network size (almost by $90\%$), but it
ensures that each user can be both source and destination of information
content.  In this way, information transfer is in principle possible between any
pair of users, and it is possible to single out the impact of the network's
dynamics. 
In Table \ref{tab:pcn} we present a summary of the main topological properties
of the PC network and its SCC.  See Methods and SI for a detailed explanation of
the data set collection.

\begin{table}[tb]
  \centering
  \def\arraystretch{1.5}
  \caption{Main properties of aggregated PC network and its {largest}
    strongly connected component (SCC): number of users $N$, with overall
    positive (negative) values of political position $N_+$ ($N_-$), total number of interactions
    $W$, and average out-degree $\av{k_\text{out}}$.  See Table S8 for the PC
    network obtained from different hashtag classification. }
  \label{tab:pcn}
  \begin{tabular*}{\linewidth}{@{\extracolsep{\fill}}c|ccccccc}
   & $N$ &  $N_+$ & $N_-$ & $W$  &	$\av{k_\text{out}}$   \\\hline
    \textit{Whole} & $285670$   & $101250$ & $125591$ & $2722504$	  &  $5.94$ 	  \\
     \textit{SCC}  & 31412 & 13925 & 16257  &  $1552389$ & 26.5   \\ \hline
  \end{tabular*}
\end{table}

Tweets carry different sentiments, that can be characterized by the hashtags
used. We assign to each tweet $t$ a sentiment, $s_t = \{-1, 0, +1\}$,
corresponding to a pro-impeachment, neutral, or anti-impeachment
sentiment, respectively, the second one meaning that a hashtag can be
used in the other two contexts. For a given user $i$, that has sent
a number $a_i$ of tweets (defined as his/her activity), we can associate a
time-ordered set of sentiments $\mathcal{S}_i = \{ s_1, s_2, \ldots, s_{a_i-1},
s_{a_i} \}$, and define his/her average sentiment, or political position $P_i$,
as
\begin{equation}
  P_i \equiv \frac{\sum_{t=1}^{a_i} s_t}{a_i},
  \label{eq:pol}
\end{equation}
which is bounded in the interval $[-1, +1]$.  This definition permits to
characterize user's political position as a continuous variable, allowing to
discern different degrees of orientation, in opposition to most common binary
measures. Since such definition  crucially depends on the hashtag
classification, we checked the robustness of our results by reconstructing also
a PC network based on a different classification of neutral hashtags.  See
Methods and SI for details.

\begin{figure*}[tb]
  \centering
  \includegraphics[width=0.99\linewidth]{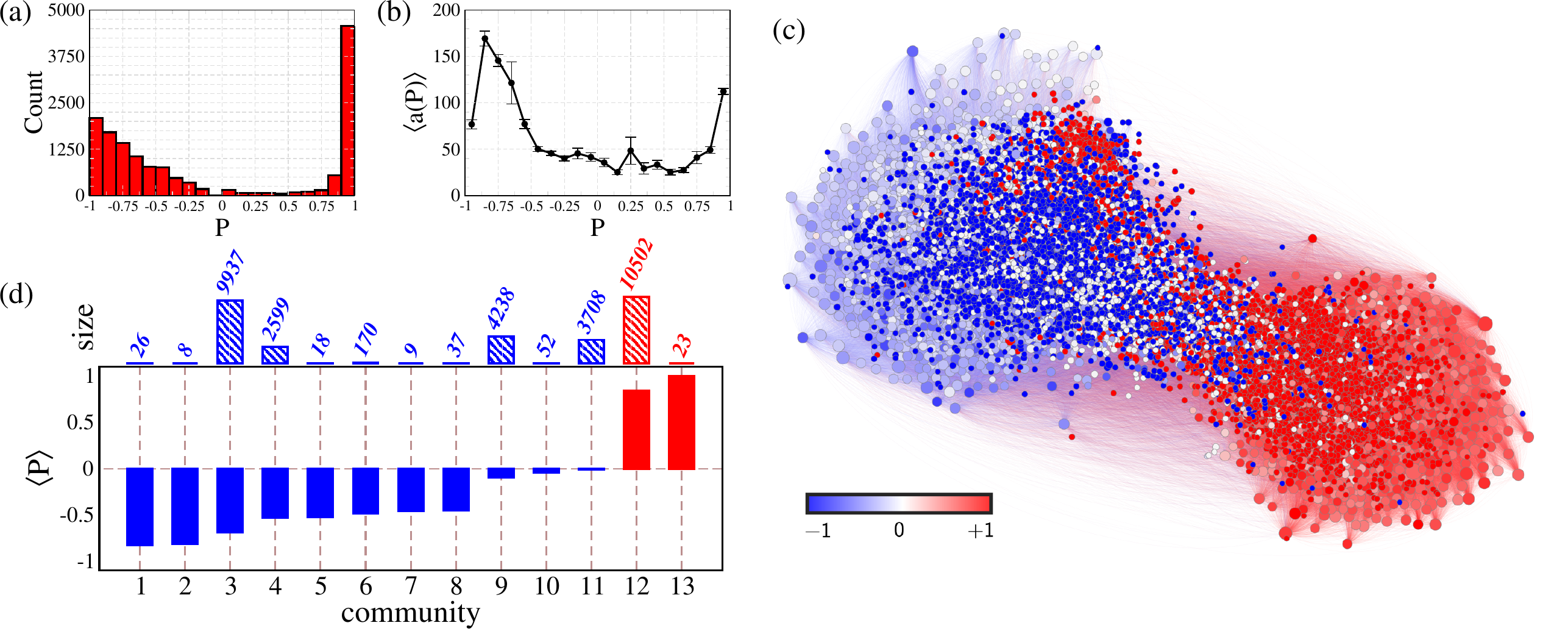}
  \caption{(a) Number of users as a function of the political position $P$. (b) Average
    activity as a function of $P$. Only users with activity {$a \ge 10$}
    in the SCC are considered for (a) and (b).  (c) Visualization of the
    time-aggregated representation of the PC network, formed by $N = 31~412$
    users in the SCC. The size of nodes increases (non-linearly) with their
    degree. Colors represent political position, as defined
    by~\eqref{eq:pol}, blue for pro-, red for anti-impeachment, and white for
    neutral average sentiment.  (d) Community size and average political position of
    different communities identified by the Louvain algorithm.}
\label{fig:net_viz}
\end{figure*}

In Fig.~\ref{fig:net_viz}a) we plot the distribution of users' political position,
showing that users are clearly split into two groups with opposite orientation,
while few users show neutral position ($P \sim 0$). Interestingly, this
 distribution is strongly asymmetric with respect to $P=0$: For
$P >0$ the great majority of users have extreme position $P \simeq +1$,
while for $P<0$ there is a decreasing variation, with more users having more
negative values of $P$.  The number of users with overall positive ($N_+$) and
negative ($N_-$) values of political positions are, however, similar, see Table~\ref{tab:pcn}.
The average sentiment of a user is inherently correlated with his/her activity.  In a
scenario in which users send tweets of opposite sentiments with
the same probability, the political position would be given by a binomial
distribution, and the expected average sentiment would decrease with
activity. Fig.~\ref{fig:net_viz}b) shows that the correlation between
average sentiment and activity is far from being driven by a random process: the
more active users are also the more extreme ones. Interestingly, for pro-impeachment
average sentiment, the most active users have $P \sim -0.75$, contrary to the case
of anti-impeachment average sentiment, in which activity is almost constant for
$ 0 < P < 0.5$, and growing for larger $P$.

Fig.~\ref{fig:net_viz}c) shows a visualization of the time-aggregated
representation of the PC network, in which users are color-coded according to
their average sentiment.  Two communities with opposite political positions are clearly
visible in the PC network, while users with neutral position are found more
frequently bridging the two groups. One can quantify this observation by
identifying the community structure~\cite{Fortunato201075} as obtained by means
of the Louvain algorithm~\cite{Blondel2008}.  In Fig.~\ref{fig:net_viz}d) we
plot the average sentiment and size of the different communities, showing
that the PC network is characterized by two larger communities, both with
approximately $10^4$ users and opposite position of similar absolute values,
$P_+\approx 0.82$ and $P_-\approx -0.70$. However, negatively polarized users
also form other communities of relevant sizes with more moderate position.
Users with strong anti-impeachment average sentiments essentially belong to a single
community, while moderate users form several communities with weak pro-impeachment
position. See SI For more details.

\section*{Topological evidence of echo chambers}
\label{sec:echo-chambers-time}

\begin{figure*}[tbp]	\scriptsize
  \centering
  \includegraphics[width=0.4\linewidth]{\FigPath/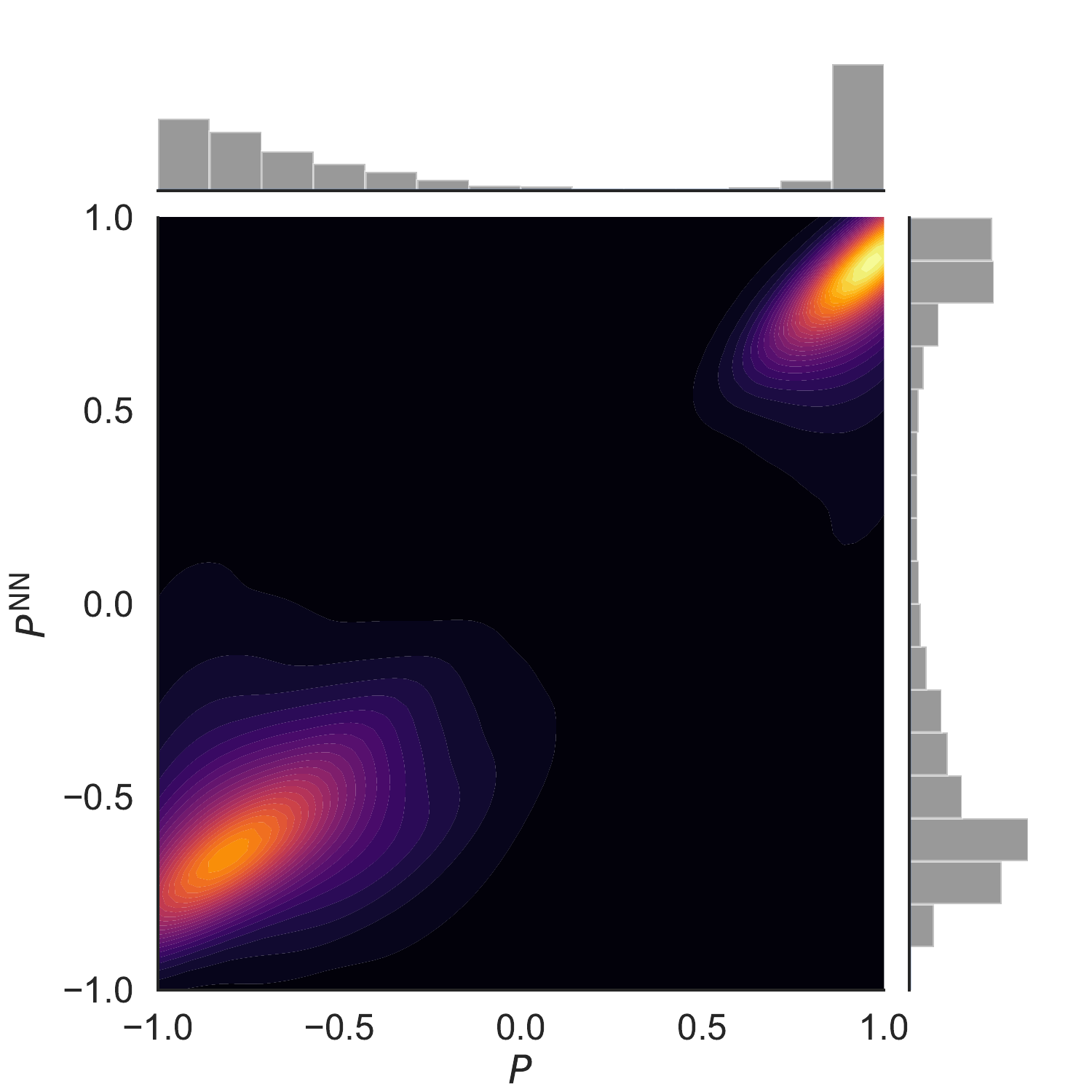}
  \includegraphics[width=0.4\linewidth]{\FigPath/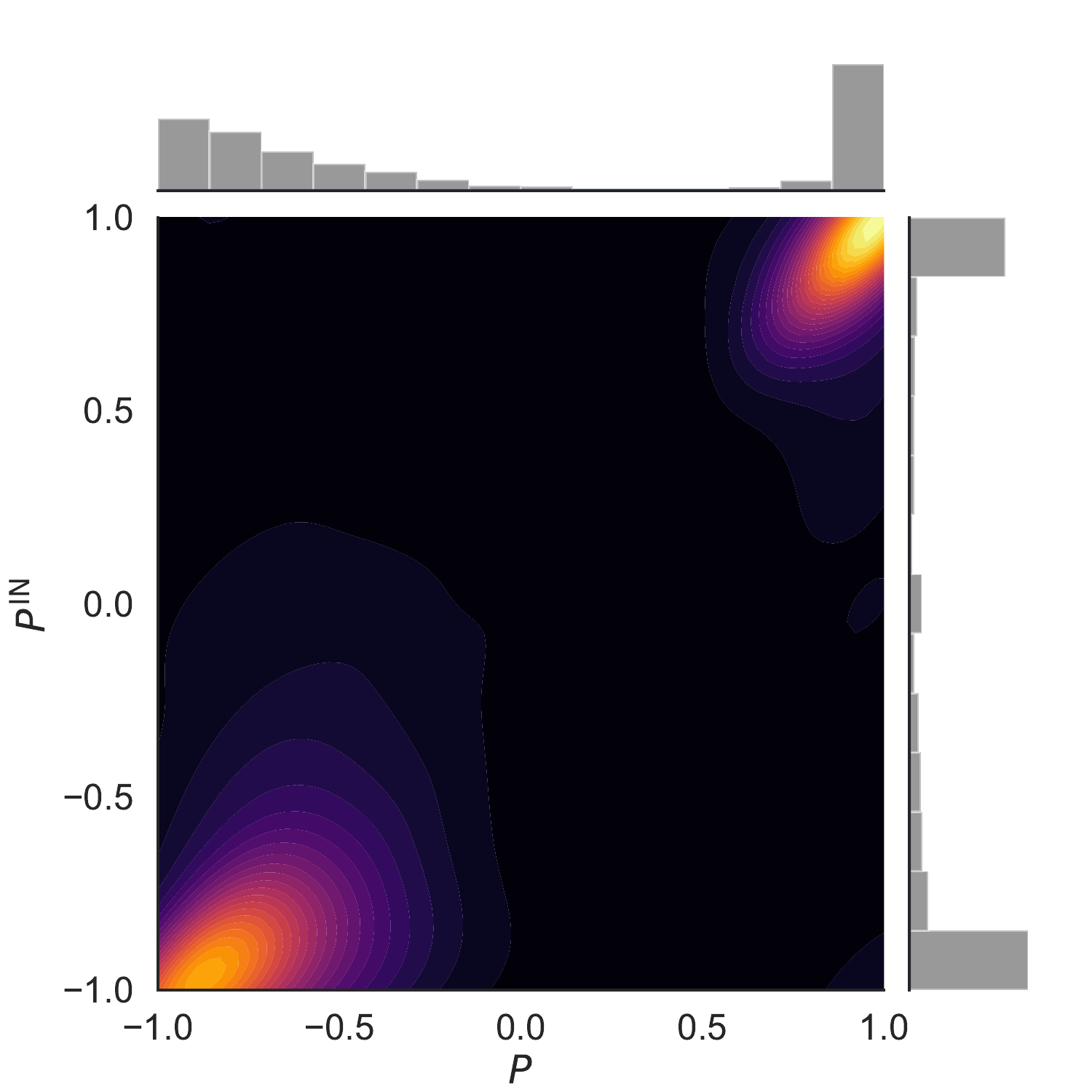}
  \caption{Contour maps for the (a) average political position of the nearest-neighbor
    $P^\text{NN}$ and (b) average sentiment of tweets received, $P^\text{IN}$
    against the average political position of a user $P$. Colors represent the density
    of users: the lighter the larger the number of users. Probability
    distribution of $P$, $P^\text{NN}$, and $P^\text{IN}$ are plotted in the
    axes. Only users with activity $a\ge 10$ (corresponding to 14813 users) are
    considered.}
  \label{fig:pol_in_out}
\end{figure*}

One can quantify the presence of echo chambers by relating the political position of a
user with the sentiment of the tweets he/she receives, as well as with the
position of his/her neighbors.  In politics, echo chambers are characterized
by users sharing similar opinions and exchanging messages with similar political
views \cite{garrett2009echo}.  This translates, at the topological level, into a
node $i$ with a given position $P_i$ connected with nodes with a
position close to $P_i$, and receiving with higher probability messages with
similar average sentiment $P_i$.  In order to quantify these insights, we define, for
each user $i$, the average position of incoming tweets, $P_i^\text{IN}$, by
applying \eqref{eq:pol} to the set of tweets from any user $j \neq i$ mentioning
user $i$.  Analogously, the average position of the nearest neighbors, or
successors, of user $i$, $P_i^\text{NN}$, can be defined as
$P_i^\text{NN} \equiv \sum_{j} A_{ij} P_j / k_{\mathrm{out}, i}$, where $A_{ij}$
is the adjacency matrix of the integrated PC network, $A_{ij}=1$ if there is a
link from node $i$ to node $j$, $A_{ij}=0$ otherwise, and
$k_{\mathrm{out}, i} = \sum_jA_{ij}$ is the out-degree of node $i$.

Figure~\ref{fig:pol_in_out} shows the correlation between the political position of a
user $i$ and (a) the position of his/her nearest neighbors, $P_i^\text{NN}$,
and (b) the average sentiment of received tweets, $P_i^\text{IN}$.  Both
plots are color-coded {contour} maps, representing the number of users in the
phase space $(P,P^\text{NN})$ or $(P,P^\text{IN})$: the lighter the area in the
map, the larger the density of users in that area.  Fig.~\ref{fig:pol_in_out}
shows a strong correlation between the position of a user and the average
position of both his/her nearest neighbors and the received tweets.  Similar
results are found for Fig.~\ref{fig:pol_in_out}(a) when considering predecessors
as nearest neighbors, 
see Fig.~S10 in the SI.  The Pearson correlation coefficient is $r=0.89$ for
$(P,P^\text{NN})$ and $r=0.80$ for $(P,P^\text{IN})$, both statistically
significant with a p-value $p < 10^{-6}$.  These topological properties of the
PC network confirm the presence of echo chambers: users expressing both pro- and
anti-impeachment are more likely to send/receive messages to/from users that
share their political opinion.

Figure~\ref{fig:pol_in_out}, however, also reveals that the densities in both
plots are not symmetric between anti- and pro-impeachment positions: for $P > 0$,
most users are concentrated in a small region of the $(P,P^\text{NN})$ and
$(P,P^\text{IN})$ spaces, while for $P<0$, they spread on a larger area.  This
means that users with extreme position $P \simeq 1$ are more likely to
interact only with users that share the same extreme position, while users
with $P<0$ exchange information (send and receive tweets) also with peers that
do not share their political opinion.  These observations are also in consonance
with the characterization of the community structure, as shown in
Fig.~\ref{fig:net_viz}(d), in which users with strong anti-impeachment average sentiment form
a single, large community, and users with pro-impeachment average sentiment form several
more heterogeneous communities.

The differences in the topological structure of the two communities can be
related to the political context under study: while users characterized by
anti-impeachment sentiments refer to a more homogeneous political area
(\textit{Partido dos Trabalhadores} and small left-wing parties),
pro-impeachment users share different political views, including center and
right-wing positions, and show different levels of sympathy in favor to the
impeachment.  Another possible and important source of asymmetry is the constant
release of content from other sources, in particular from the traditional media,
broadcasting mostly contents that stimulate pro-impeachment sentiments, possibly
reinforcing their dissemination to a more diversified audience.

\section*{Effects of political position on information spreading}
\label{sec:spreadability}

The presence of echo chambers implies that users mainly exchange messages with
other users sharing similar sentiments. This fact can have an impact on the way
in which information is transmitted through the whole PC networks. A possible
empirical way to gauge the effects of echo chambers on information spreading is
to consider the number of RTs that a given user can
achieve~\cite{Galuba:2010:OTP:1863190.1863193,Jenders2013,Ratkiewicz2011}. One
can expect that more influential users, producing content that attracts more
interest, will be rewarded by a larger number of RTs. Figure~S11 (see SI) shows
the number of RTs of the users, as a function of both his/her activity and
position. One can see that the number of times that a user is re-tweeted is
strongly correlated with the activity of that user. With hindsight, this
observation is to be expected, since a user that produces many tweets gets a
larger chance of being re-tweeted, within a homogeneous assumption of equal
probability of re-tweeting. However, if we consider the number of RTs normalized
by the total tweets sent, we observe a lack of evident correlations with users'
political position, as shown in Fig.~S11(b) in SI.

In order to better understand the role of the network's polarization in
information propagation, we followed a different approach, by considering simple
models of spreading dynamics. We have focused  in the
susceptible-infected-susceptible (SIS) and susceptible-infected-recovered (SIR)
models~\cite{anderson92}, classical epidemic processes which have also been used to
study the diffusion of information~\cite{ZHAO2013995,PhysRevLett.111.128701}. 
In the SIS model, each agent can be in either of two states, susceptible or
infectious, whereas in SIR it can also be in a recovered state in which it
cannot be infected nor transmit the disease. Susceptible agents may become
{infectious} upon contact with infected neighbors, with certain transmission
rate $\lambda$ in both processes. Infectious agents can spontaneously heal with
rate $\tau^{-1}$, becoming susceptible again or recovered in SIS and SIR,
respectively.  Within an information diffusion framework, a susceptible node
represents a user who is unaware of the circulating information (e.g.~rumors,
news, an ongoing street protest), while an infectious user is aware of it and
can spread it further to his contacts. A recovered agent is aware but not
willing to transmit the information.

We ran the SIS and SIR dynamics on the temporal PC network, using the real
timing of connections between users as given by the time stamps of interactions,
so to ensure that the information diffusion follows time-respecting paths. In
temporal networks characterized by an instantaneous duration of contacts, the
infection process can be implemented by considering $\lambda$ as a transmission
probability, i.e. whenever a susceptible node $i$ gets in contact with an
infectious node $j$, node $i$ will become infected with probability $\lambda$.
The healing occurs spontaneously after a fixed time $\tau$ with
respect to the moment of infection. We start the dynamics with only one node $i$
infected, and stop it on the last interaction of the temporal sequence. The set
of nodes that were infected at least once along the dynamics, started with $i$ as
source of infection, forms the \textit{set of influence} of node $i$,
$\mathcal{I}_i$~\cite{PhysRevE.71.046119}. The set of influence of a user thus
represents the set of individuals that can be reached by a message sent by
him{/her}, depending on the transmission probability $\lambda$ and healing time
$\tau$.

\begin{figure}[tbp]\centering
  \includegraphics[width=0.99\linewidth]{\FigPath/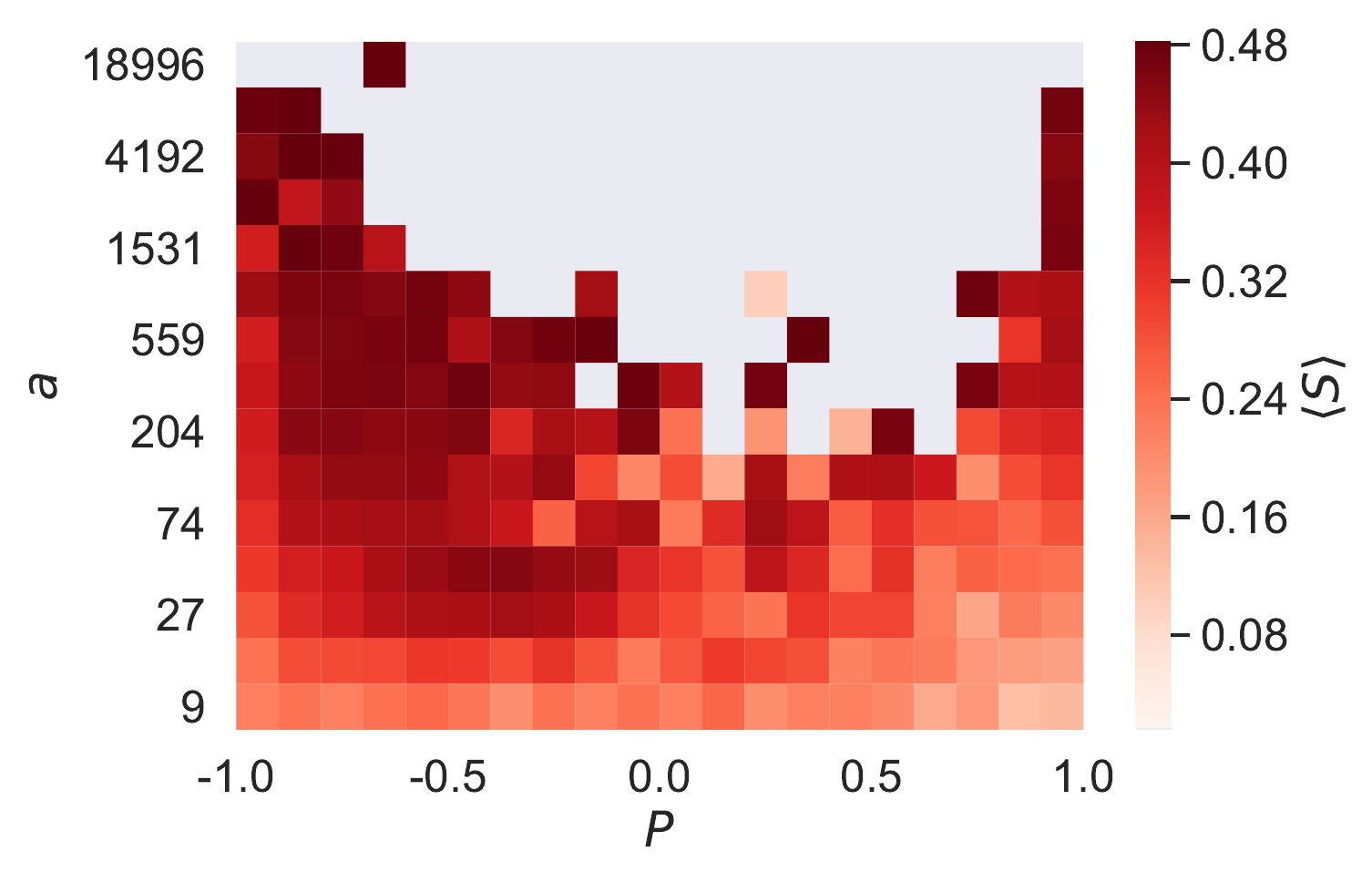}
  \caption{Heat map of the average spreadability $\av{S}$ of users, as a
    function of their political position $P$ and activity $a$.  The transmission
    probability of the {SIS} dynamics is {$\lambda = 0.5$ and $\tau = 7$ days}.
    Averages were performed over 100 runs.}
  \label{fig:heatmaps}
\end{figure}

For different values of $\lambda$ and $\tau$, we measure the
\textit{spreadability} $S_i$ of each user $i$, defined as the relative size of
his/her set of influence, namely
\begin{equation}
  \label{eq:spreadabiltiy}
  S_i(\lambda, \tau) \equiv \frac{|\mathcal{I}_i(\lambda, \tau)|}{N},
\end{equation}
by running a SIS or SIR dynamics with node $i$ as seed of the infection,
averaged over several runs.  In Fig.~\ref{fig:heatmaps} we plot the average
spreadability $\av{S}$ of users as a function of their political position $P$
and activity $a$ for the SIS model.  As expected, the more active are the users,
the larger their spreadability (darker colors of the plots). However, one can
see that $\av{S}$ {is} not constant with respect to the users' political
orientation: the spreadability is clearly smaller for users with
anti-impeachment average sentiment, while it is larger for users with $P<0$,
reaching a maximum for $P \sim -0.5$. Different values of $\lambda$ and $\tau$
for the SIS and SIR model (available in the SI) show similar behavior.

In order to disentangle the effect of the users' political position on
spreadability from the effect of activity, in Fig.~\ref{fig:spread_pol} we plot
the average spreadability of users as a function of their position, $\av{S(P)}$,
for $\lambda=0.2$ and $\tau=7$ days. Other values are shown in the SI.  Only users
with activity bounded {in the interval} $a \in [10,100]$ {are considered}, so as
to ensure that the average users' activity is relatively homogeneous with
respect to their political position (as shown in the SI).  Figure
\ref{fig:spread_pol} shows that the average spreadability reaches a maximum for
users with intermediate pro-impeachment position, $P \simeq -0.5$, maximum that
is up to four times larger than the value for users with anti-impeachment position. 
This striking difference is robust with respect to the value of the transmission
probability $\lambda$ and healing time $\tau$. As shown in the SI, the shapes of
the $\av{S(P)}$ curves are remarkably similar, even though significantly
different values are reached. Analogous behavior is observed for the SIR model
(see SI).

\section*{Diversity increases spreadability}
\label{sec:diversity}

The origin of the large spreadability of  users with pro-impeachment position cannot be
traced back to their numeric prevalence in the network, since users are split
into two groups of {similar} size; see Table~S8 in SI.  Moreover, the great
majority of users are characterized by extreme position, $|P| \simeq 1$, yet
they show a much smaller spreadability than users with intermediate pro-impeachment
position, $P \simeq -0.5$.  One way to understand this difference relies in
looking at the characteristics of the users reached by the spreading dynamics.
One can analyze the political position of the set of influence $\mathcal{I}_i$, by
defining, for each user $i$, the average $\mu_i$ and the variance $\sigma_i$ of
the political positions expressed by $\mathcal{I}_i$, as
\begin{equation}
  \mu_{i}\equiv  \sum_{j \in \mathcal{I}_i} \frac{ P_j}{|\mathcal{I}_i|},    \qquad
  \sigma_i \equiv \sum_{j \in \mathcal{I}_i} \frac{(P_j -
    \mu_i)^2}{|\mathcal{I}_i|}.
\end{equation}
The average  $\mu_i$ represents the average opinion of the users reached by $i$,
while the variance $\sigma_i$ represents how heterogeneously oriented
$\mathcal{I}_i$ is.  A small variance $\sigma_i$ indicates that the political
position of $\mathcal{I}_i$ is quite uniform and close its average value, while
a large value of $\sigma_i$ shows that $\mathcal{I}_i$ is heterogeneous with
respect to the political position.  Therefore, the variance $\sigma_i$
quantifies the \textit{diversity} of the users reached by $i$.

\begin{figure}[tb]
  \begin{center}
    \includegraphics[width=0.99\linewidth]{\FigPath/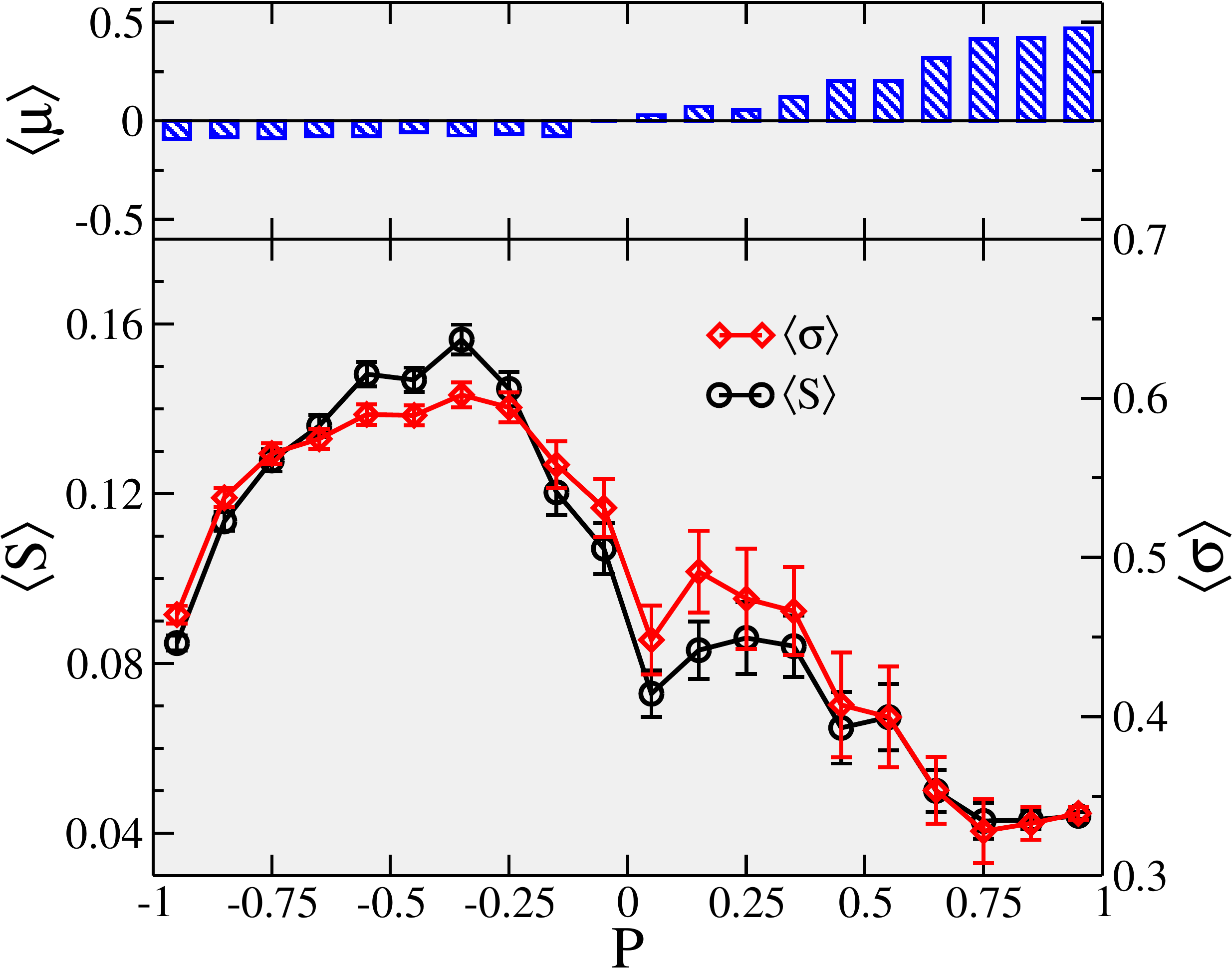}
  \end{center}
  \caption{Average spreadability $\av{S (P)}$ (black curve, left axes) of users
  	with political position $P$.  Average diversity $\av{\sigma(P)}$ ({red curve,
  		right axes}) and political position $\av{\mu(P)}$ ({bars, top panel}) of the
  	set of influence reached by users with position $P$.  Transmission probability
  	{$\lambda = 0.20$ and $\tau = 7$~days}.  Only {the 11386} users with activity
  	$a \in [10,100]$ are considered.  Different {ranges of $a$ and} values of
  	$\lambda$ are shown in the SI.  Results are averaged over 100 runs, error bars
  	represent the standard error.}
  \label{fig:spread_pol}
\end{figure}

In Fig. \ref{fig:spread_pol} (top panel) we plot the average political position
$\av{\mu(P)}$ of the set of influence reached by users with position $P$,
showing that users with pro-impeachment (neutral, anti-impeachment) average
sentiment are more likely to reach, on average, users sharing the same
pro-impeachment (neutral, anti-impeachment) average sentiment.  This result
(robust across different values of $\lambda$ and $\tau$, as shown in the SI)
indicates that, given the strongly polarized structure of the network,
information diffusion is biased toward individuals that share the same political
opinion, quantifying the effect of echo chambers. The average $\av{\mu(P)}$,
indeed, gauges the strength of the echo chambers: the more $\av{\mu(P)}$ is
close to $P$, the stronger the echo chamber effect. Furthermore, one can note
differences between  users with pro- and anti-impeachment positions, $\mu$ is
almost constant for negative values of $P$, so echo chamber effects are small,
while $\mu$ is growing almost linearly for positive $P$, indicating stronger
echo chambers effects.

Even more interesting, Fig. \ref{fig:spread_pol} shows that the diversity
$\sigma_i$ of the users reached by $i$ strongly depends on his{/her} political position
$P_i$. The curve of the average diversity as a function of the position,
{$\av{\sigma(P)}$}, follows a behavior remarkably similar to the average
spreadability of users with position $P$, $\av{S(P)}$. The strict correlation
observed between {$\av{\sigma(P)}$} and $\av{S(P)}$ indicates that if a user is
able to reach a diverse audience, formed by users that do not share his average
sentiment, then the size of his/her set of influence is much larger. That is,
individuals with large spreadability are able to break their echo chambers. 
Note that this result is not trivial since the size of the echo chambers are
much bigger than the number of users reached. Moreover, the value of
$\av{\sigma(P)}$ is statistically significant and does not depend on the number
of users considered in the average. For instance, there are much more users with
extreme orientations  ($|P| \simeq 1$) than users with intermediate position
($P \simeq -0.5$), yet it holds $\av{\sigma(P \simeq -0.5 )} \gg \av{\sigma(|P|
	\simeq 1)}$. Furthermore, given the larger number of users considered, error
bars for $\av{\sigma(|P| \simeq 1)}$ are smaller than the ones for $\av{\sigma(P
	\simeq -0.5 )}$.

%
%

\section*{Discussion}
\label{sec:discussion}

The effects of echo chambers on the openness of online political debate have
been argued by the scientific community. Recently, it has been shown that echo
chambers are expected to enhance the spreading of information in synthetic
networks \cite{10.1371/journal.pone.0203958}. Their impact in real communication
networks, however, remains poorly understood.  The main contribution of this
paper is twofold: i) we quantify the presence of echo chambers in the Twitter
discussion about the impeachment of former Brazilian President Dilma Rousseff,
showing that communities of users expressing opposite political positions
emerge in the topological structure of the communication network, and ii) we
gauge the effects of such echo chambers on information spreading, showing that
the capability of users to spread the content they produce depends on their
political orientation. The use of spreading models allows us to characterize the
internal structure of echo-chambers, showing that users belonging to the same
echo chamber, with different convictions (i.e., the intensity of their attitude
to the impeachment issue), can have quite different spreading capabilities.

Our method to quantify echo chambers is built upon two main ingredients: i) we
reconstruct a communication network based in mentions, in which people can
actually discuss and exchange information related to ongoing political events,
and ii) we define a continuous political position measure, by classifying the hashtags
used in tweets as expressing a sentiment in favor or against the impeachment,
which is independent by the network's reconstruction. We then observe that anti-
and pro-impeachment sentiments clearly separate into different communities in
the PC network. It is important to remark that, while it is well known that
networks formed by RTs can be have a strong partisan structure, since RTs
generally imply endorsement, this observation is new for mention networks, in
which users characterized by opposite average sentiments can easily interact
\cite{Truthy_icwsm2011politics}.

These two clusters of users sharing similar opinions, or echo chambers, can be
characterized by looking at the correlations between the in-flow and out-flow of
sentiments, as well as between the average sentiments of an individual and his/her
nearest neighbors.  The topologies of the two echo chambers, however, are not
exactly equivalent.
Users expressing anti-impeachment sentiments tend to lean towards the extreme,
achieving a position $P \simeq +1$, while users with pro-impeachment
sentiments show smoother tendencies, reflected into the presence of medium-sized
communities with overall negative position.

We have gauged the effects of echo chambers on information diffusion by running
simple models of information spreading, observing that, on average, users are
more likely to receive information from peers sharing the same average
sentiments. We then see that people with predominantly pro-impeachment 
sentiments are able to broadcast their message to a potentially larger audience
than other users. Furthermore, such audiences are also characterized by a
greater diversity of opinions, indicating that pro-impeachment sentiments can
spread to users expressing both  pro- and anti-impeachment attitudes, a
signature that echo chambers can be broken. An interesting question arising here
is what makes pro-impeachment users better spreaders than users with an opposite
view. Recent works~\cite{Banos2013,Alvarez2015} have related spreading
efficiency of users with their topological position  in the integrated network,
in particular with the degree and centrality of individuals as measured by the
$k$-core index~\cite{Seidman1983269}. In Fig.~S16 of the SI we show that the
average position $P$ of users is quite uncorrelated with both their average
degree $k$ and $k$-core index, indicating that users characterized by
pro-impeachment average sentiment cannot be single out on simple topological
basis.

It is important to highlight that our method for quantifying the echo-chamber
effects by using epidemic processes comes at the cost of limitations.  A first
issue is that only very large communication networks can be analyzed, due to the
extraction of the strongly connected component that greatly reduces the number
of nodes. However, this procedure is essential to properly address the
communication dynamics between users, and possibly avoid the presence of social
bots.  Furthermore, our definition of political position entirely relies on the hand
tagged hashtags classification.  It is well known that hashtags can be
hijacked~\cite{Hadgu:2013:PHH:2487788.2487809}, i.e.~they can be used by some
users with a different (or opposite) purpose than the one originally intended,
thus invalidating the sentiment inferred through it.  However, our analysis is
based on a large number of hashtags, and it is robust with respect to a
significant change of the sentiment classification; see results for the
additional classification in the SI.

Future research in this topic should address three main points. Firstly, more
sophisticated methods for detecting users' political position, such as automatic
sentiment analysis of tweet contents could be considered. These methods are,
however, not exempt from
limitations~\cite{Goncalves:2013:CCS:2512938.2512951,Conover2010predicting}.
Secondly, more realistic models of information diffusion, such as complex
contagions, independent cascade and linear threshold
models~\cite{Watts5766,Centola,Saito:2008:PID:1430307.1430318,Borodin:2010:TMC:1940179.1940229}, could be used to estimate individual's spreadability. We have checked numerically that a modification  of the classic Watts threshold model for complex contagion~\cite{Watts5766} to the framework of temporal networks~\cite{karimi2013} leads to the same behavior observed in the SIR and SIS models, see Fig. S17 of the SI.  Therefore, while we do not expect our results to qualitatively depend on the details of propagation dynamics considered, interesting features may be added, such as a transmission probability that depends on the similarity between opinions. It would also be interesting to measure the evolution of users' political position in time, as they are expected  to not be constant over the whole temporal sequence.  Finally, given that our conclusions are based in a single case study, it would be interesting to replicate our method in different data sets, to identify in a quantitative way the presence of echo chambers across differently polarized political contexts, over different social media.

\section*{Methods}
Here we describe the empirical data used in the paper,
  available upon motivated request to the authors, and how we
  reconstruct the network from it, as well as the results of the
  hashtags classification. For further details, see SI.

\subsection*{Reconstruction of the PC networks}

Our data set is composed of tweets collected daily from the public
streaming of the Twitter API by specifying a list of 323 keywords (See
Table~S2 of SI) related to the impeachment process of former president
of Brazil, Dilma Rousseff. Data have been gathered between March 5th to
December 31st of 2016.
Only tweets including mentions to other users and at least one of the classified
hashtags (see next Section) have been selected, while retweets have been
discarded.  Tweets containing hashtags of opposite sentiments ($s_t=+1$ and
$s_t=-1$) are less than $1\%$, and have been discarded.  The timing of the
interactions has been preserved, so that in the temporal PC network a directed
link from node $i$ to node $j$ at time $t$ is drawn if user $i$ sends a tweet by
mentioning $j$ at time $t$.  Finally, the strong component of the
time-aggregated version of the PC network has been extracted.

\subsection*{Hashtag classification}

A list of the 495 most tweeted hashtags from the collected data has been
classified by performing a manual annotation of the sentiments
(anti-, pro-impeachment, neutral, or not related to the issue) by four
independent volunteers.  Through an interactive webpage, the volunteers
had the opportunity to browse Twitter for checking tweets containing the
selected hashtag within the time window of interest.  The final
classification of each hashtag has been determined by the majority (3 of 4)
of the opinions of the volunteers.  A number of 321 ($64.8\%$) hashtags
had a full agreement, while in 443 ($89.5\%$) of them at least 3 of 4
persons agreed. A majority agreement has not been reached for 52 ($10.5\%$)
hashtags, which have been excluded from the data set.  Discrepancies
between any pair of volunteers were less than $10\%$.  A final list of
404 hashtags (see Table S3 {to S6} in the SI for final classification) has been
used to reconstruct the PC network.

\text{ }\\
\noindent  \textbf{Acknowledgments:}
We thank Gino Ceotto and Diogo H. Silva for voluntarily performing  the hashtag
classification. This work was partially supported by the Brazilian agencies CNPq
and FAPEMIG. Authors thank the support from the program \textit{Ci\^encia sem
	Fronteiras} - CAPES under project No. 88881.030375/2013-01. This study was
financed in part by the Coordena\c{c}\~{a}o de Aperfei\c{c}oamento de Pessoal de
N\'{\i}vel Superior - Brasil (CAPES) - Finance Code 001. M.S. acknowledges
financial support by the J. McDonnell Foundation.  R.P.-S. acknowledges
financial support from the Spanish MINECO, under Project No.
FIS2016-76830-C2-1-P, and additional financial support from ICREA Academia,
funded by the Generalitat de Catalunya.

\par
\text{ }\\
\noindent  \textbf{Author contributions:}
W.C., S.C.F., R.P.-S., and M.S. designed and performed research. W.C. collected data and performed simulations.  W.C., S.C.F., R.P.-S., and M.S. wrote the paper.

\par
\text{ }\\
\noindent  \textbf{Additional Information:} The authors declare no competing  interests.

\par
\text{ }\\
\bibliography{refs_Twitter_EC}

\end{document}